\documentclass[twocolumn,english]{revtex4-2}
\usepackage[T1]{fontenc}
\usepackage[latin9]{inputenc}
\usepackage{babel}
\usepackage[unicode=true,pdfusetitle,
 bookmarks=true,bookmarksnumbered=false,bookmarksopen=false,
 breaklinks=false,pdfborder={0 0 1},backref=false,colorlinks=false]
           {hyperref}
\usepackage{amsmath,amssymb}
\usepackage{graphicx}
\usepackage{enumerate}
\usepackage{natbib}
\usepackage{bm}
\usepackage{amsfonts}
\usepackage{babel}
\usepackage{array}
\usepackage{url} 
\usepackage{mathtools}

\def\xsum{\mathop{\sum_k\nolimits'}}
\def\msum{\mathop{\sum_{k,k\ne\,m}\nolimits'}}

\begin{document}
\title{Inverse Sampling of Degenerate Datasets from a Linear Regression Line}
\author{Albert S. Kim}
\affiliation{Civil and Environmental Engineering, University of Hawaii at Manoa}
\date{\today}
\email{albertsk@hawaii.edu}

\homepage{http://albertsk.org}

\begin{abstract}
  When linear regression generates a relationship between a (dependent)
scalar response and one or multiple independent variables, various
datasets providing distinct graphical trends can develop resembling
relationships based on the same statistical properties. Advanced statistical
approaches, such as neural networks and machine learning methods,
are of great necessity to process, characterize, and analyze these
degenerate datasets. On the other hand, the accurate creation of purposedly
degenerate datasets is essential to test new models in the research
and education of applied statistics. In this light, the present study
characterizes the famous Anscombe datasets and provides a general
algorithm for creating multiple paired datasets of identical statistical
properties. 
\end{abstract}
\maketitle

\section{Introduction}
\label{sec:intro}
Originally termed the least-squares fitting, the linear regression
method is one of the most widely used analysis tools to primarily
investigate trends among variables in various disciplines.
Legendre and Gauss initially formulated the regression method from
the late 18th to early 19th centuries to understand observed datasets
of astronomical phenomena. The modern statistical characteristics
of the regression were initially established by Galton's work that
described biological phenomena \citep{massey_kolmogorov-smirnov_1951,galton_kinship_1989,ireland_effect_2016},
followed by Yule \citep{yule_theory_1897}'s and Pearson \citep{pearson_law_1903}'s
early mathematical formulation. When a linear relationship of a paired dataset
provides two fitting coefficients, i.e., the intercept and the slope,
the goodness of the regression is often evaluated by the coefficient
of determination, denoted as $R^{2}$. Although these three outputs
provide a good understanding of how the independent variable $x$ is
quantitatively correlated to the response variable $y$, the linear
regression's inherent problem resides in its statistical degeneracy,
such that multiple datasets can have indistinguishable statistical
properties.

A quartet of visually distinct graphs, having identical regression
statistics, were investigated by Anscombe \citep{anscombe_graphs_1973},
who emphasized  the equal significances of graphical visualization
and quantitative statistics  \citep{cook_graphs_1999,rousselet_beyond_2017}.
The noticeable heterogeneity of his work's graph patterns conversely
emphasizes the significance of the data degeneracy  \citep{cook_detection_1977}.
Nevertheless, his data generation method was only partially studied \citep{murray_generating_2021}, and to the best of our knowledge,
the full mechanism is still unknown  \citep{schneider_adding_2013,murray_generating_2021}, even if each dataset
has only 11 pairs.

In principle, simple linear regression between two variables can be easily extended
to multiple and non-linear regressions, which include several variables
and their power-wise products, respectively. Regardless of the regression
type, a regression method uses a single matrix to relate the input(s)
and output(s), and the matrix elements consist of, in general, various
products of input variables. To investigate relationships between
highly correlated data, multiple matrices can be inserted between the
input and output layers, and their elements can be calculated using
various non-linear functions \citep{smith_mutual_2015}. 
Neural networks and machine learning  \citep{hosseinzadeh_application_2020,khademi_multiple_2017,lin_improving_2009} are some advanced methods within a category of data exploration \citep{shoresh_data_2012}.

Once a relationship is made, as either an empirical equation or a
matrix form, the range of input variables often limits the applicability
of the regression, leaving infinite degrees of degeneracy. There can
possibly be many combinations of input variables that provide the
same output results. For both preliminary tests of any new, advanced
regression algorithm, it is necessary to have a data generator that
can create manyfold datasets, satisfying the same statistical constraints.
In this light, this work revisits linear regression fundamentals,
analyzes Anscombe's quartet data, and provides a possible algorithm
to inversely create degenerate datasets of distinct values with predetermined
statistical parameters \citep{halperin_inverse_1970}.

\section{Linear Regression Theory}

We consider a linear model, such as 
\begin{equation}
\bm{y}  =\beta_{0}+\beta_{1}\bm{x}+\bm{\epsilon}
\end{equation}
where $\bm{x}=\left\{ x_{1},x_{2},\cdots,x_{N}\right\} $ and $\bm{y}=\left\{ y_{1},y_{2},\cdots,y_{N}\right\} $
are vectors of $N$ (observed) elements for the independent and response
(dependent) variables, respectively; $\bm{\epsilon}$ is a vector
of randomly distributed errors of zero mean and finite variance, and
$\beta_{0}$ and $\beta_{1}$ are regression or fitting parameters,
so called the $y$-intercept and slope, respectively. Here, we define
the regression function, such as 
\begin{equation}
\bm{Y}   =\beta_{0}+\beta_{1}\bm{x}\label{eq:LR}
\end{equation}
that most closely fits the paired data of $\left(\bm{x},\bm{y}\right)$
of size $N$. Here, statistically meaningful properties include the
mean and variance of $\bm{x}$, i.e., $\bar{x}=\text{mean}\left(\bm{x}\right)$
and $\sigma_{x}^{2}=\text{var}\left(\bm{x}\right)$, respectively;
those of $\bm{y}$, i.e., $\bar{y}=\text{mean}\left(\bm{y}\right)$
and $\sigma_{y}^{2}=\text{var}\left(\bm{y}\right)$, respectively;
and the parameter $\beta_{1}$ for the $N$ paired points. The goodness
of the regression is estimated using the coefficient of determination,
denoted as $R^{2}$, defined as 
\begin{equation}
R^{2}=\frac{\sum_{k}\left(Y_{k}-\bar{y}\right)^{2}}{\sum_{k}\left(y_{k}-\bar{y}\right)^{2}}=\beta_{1}^{2}\frac{S_{xx}}{S_{yy}}=\beta_{1}^{2}\frac{\sigma_{x}^{2}}{\sigma_{y}^{2}}\label{eq:defR2}
\end{equation}
and 
\begin{equation}
\beta_{1}=\frac{S_{xy}}{S_{xx}}=\frac{\sigma_{xy}}{\sigma_{x}^{2}}\label{eq:def-beta1}
\end{equation}
where $\sigma_{xy}$ is a covariance between $\bm{x}$ and $\bm{y}$;
$S_{xx}$ and $S_{yy}$ are sums of squares of residuals, i.e., $\sum_{k}\left(x_{k}-\bar{x}\right)^{2}$
and $\sum_{k}\left(y_{k}-\bar{y}\right)^{2}$, respectively; and $S_{xy}$
is a sum of residual products, i.e., $\sum_{k}\left(x_{k}-\bar{x}\right)\left(y_{k}-\bar{y}\right)$.
The magnitude and the sign of $\beta_{1}$ are given as those of $R\sigma_{y}/\sigma_{x}$
and $S_{xy}$, respectively. Given a paired dataset, the linear regression
process indicates the calculation of $\beta_{1}$ and $\beta_{0}$
values that minimize the error $\bm{\epsilon}$, and is often straightforward,
using various spreadsheet programs or numerical/statistical packages,
such as Microsoft Excel, Google Sheets, MATLAB/Octave, python, and
R-language. In applied statistics disciplines, it is also important
to generate manyfold datasets that accurately satisfy the predetermined
statistical properties for various testing and training purposes.

\subsection{Revisit to Anscombe's Quartet}

\begin{figure}
\begin{centering}
\includegraphics[width=3.25in]{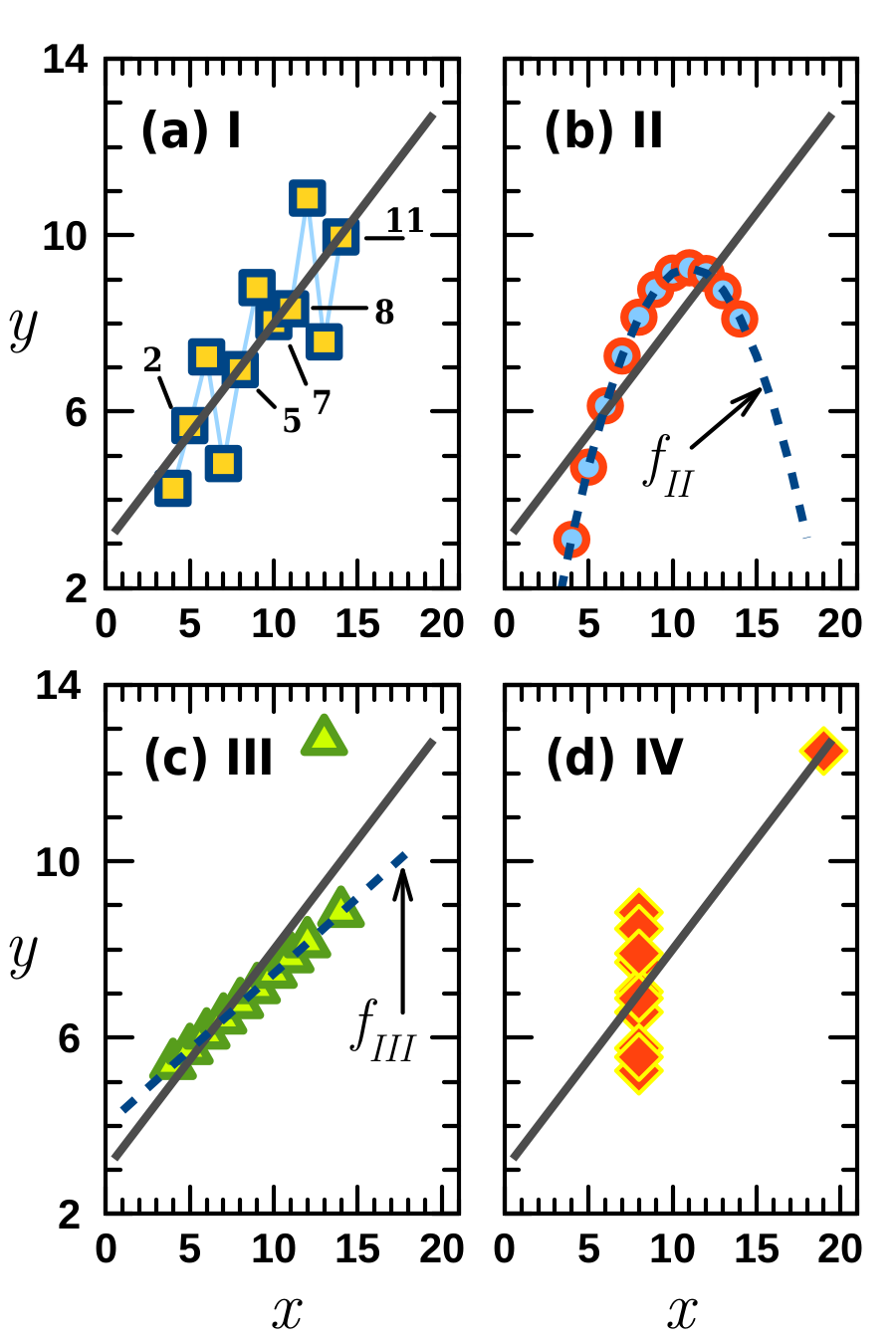}
\par\end{centering}
\caption{Plots of Anscombe's four data sets with the linear regression line
of $Y=3.0+0.50\,x$, with the shape functions discussed in section
\ref{subsec:shape-function}.}
\label{fig:Plots-of-Anscombe-Quartet}
\end{figure}

Anscombe's original quartet, i.e., datasets I--IV, listed in Table
\ref{tab:origAnscom}, is visualized in Fig. \ref{fig:Plots-of-Anscombe-Quartet},
representing graphically distinct patterns of $y$'s with respect
to $x$. A brief analysis of the quartet is as follows. Fig. \ref{fig:Plots-of-Anscombe-Quartet}(a)
shows an apparently linear trend of dataset I, typical in studies
of various disciplines. Fig. \ref{fig:Plots-of-Anscombe-Quartet}(b)
(of circular symbols) shows a parabolic, concave-down trend of $y$,
having its peak at $\left(x,y\right)=\left(11,9.26\right)$. 
Most points in dataset I and II are closely located
near the linear trend line. On the other hand, Fig. \ref{fig:Plots-of-Anscombe-Quartet}(c)
has a noticeable outlier above a linear line that passes through the
vicinity of the rest of the 10 points. Fig. \ref{fig:Plots-of-Anscombe-Quartet}(d)
has a bimodal distribution of the 11 data points, i.e., a group of
10 points at one $x$-coordinate and one outlier away from the group.
  Interestingly, the four datasets of the distinct patterns contain
identical statistical properties, summarized in Table \ref{tab:PropAnscom}.
In each dataset, the sample size is equally $N=11$; the mean and
variance of $\bm{x}$ are $\bar{x}=9.0$ and $\sigma_{x}^{2}=11.00$,
respectively; and those of $\bm{y}$ are $\bar{y}=7.5$ and $\sigma_{y}^{2}=4.125$,
respectively. (In Anscombe's original work, sums of $\left(x_{k}-\bar{x}\right)^{2}$
and $\left(y_{k}-\bar{y}\right)^{2}$ are reported, instead of variances,
as 111.0 and 41.25, respectively.) The regression
statistics provide the same values of $\beta_{1}=0.5$, $\beta_{0}=3.0$,
and $R^{2}=0.667$, with acceptable errors.  To the best of our knowledge,
how Anscombe generated the data quartet has not been well explained
in the literature. Because our goal is to create multiple degenerate
datasets of the same statistical properties, we here investigate the
characteristics of Anscombe's quartet data in detail.

For a better understanding, we first sorted the datasets in Table
\ref{tab:origAnscom} in an ascending order of $\bm{x}$ and made
Table \ref{tab:SortedAnscom}. Note that the sequences of data pairs
do not change the statistical results of the linear regression. 
 For example, even if the first two points of dataset I in Table
\ref{tab:origAnscom}, i.e., $\left(x_{1},y_{1}\right)=\left(10.0,8.04\right)$
and $\left(x_{2},y_{2}\right)=\left(8.0,6.95\right)$, are exchanged
to $\left(x_{1},y_{1}\right)=\left(8.0,6.95\right)$ and $\left(x_{2},y_{2}\right)=\left(10.0,8.04\right)$,
the regression statistics of $\beta_{0}$, $\beta_{1}$, and $R^{2}$
values remain invariant.   In Table \ref{tab:SortedAnscom}, it
is noticed that datasets I--III have an evenly distributed $\bm{x}$
from 4 to 14 with a fixed interval of 1, and the outliers ($y>10$)
of datasets III and IV are located near the end of the regression
line in $x$. Now, we explain how the $\bm{x}$ and $\bm{y}$ vectors were
possibly generated, keeping the statistical constraints discussed
above.
\begin{table}
\begin{centering}
\caption{Anscombe's original quartet datasets.}
\par\end{centering}
\begin{centering}
\begin{tabular}{|c|r|r|r|r|r|r|}
\hline 
 &  & \multicolumn{1}{r|}{I} & \multicolumn{1}{r|}{II} & \multicolumn{1}{r|}{III} & \multicolumn{2}{r|}{IV}\tabularnewline
\hline 
\hline 
Index & $x$ & $y$ & $y$ & $y$ & $x$ & $y$\tabularnewline
\hline 
1 & 10.0 & 8.04 & 9.14 & 7.46 & 8.0 & 6.58\tabularnewline
\hline 
2 & 8.0 & 6.95 & 8.14 & 6.77 & 8.0 & 5.76\tabularnewline
\hline 
3 & 13.0 & 7.58 & 8.74 & 12.74 & 8.0 & 7.71\tabularnewline
\hline 
4 & 9.0 & 8.81 & 8.77 & 7.11 & 8.0 & 8.84\tabularnewline
\hline 
5 & 11.0 & 8.33 & 9.26 & 7.81 & 8.0 & 8.47\tabularnewline
\hline 
6 & 14.0 & 9.96 & 8.10 & 8.84 & 8.0 & 7.04\tabularnewline
\hline 
7 & 6.0 & 7.24 & 6.13 & 6.08 & 8.0 & 5.25\tabularnewline
\hline 
8 & 4.0 & 4.26 & 3.10 & 5.39 & 19.0 & 12.50\tabularnewline
\hline 
9 & 12.0 & 10.84 & 9.13 & 8.15 & 8.0 & 5.56\tabularnewline
\hline 
10 & 7.0 & 4.82 & 7.26 & 6.42 & 8.0 & 7.91\tabularnewline
\hline 
11 & 5.0 & 5.68 & 4.74 & 5.73 & 8.0 & 6.89\tabularnewline
\hline 
\end{tabular}
\par\end{centering}
\centering{}\label{tab:origAnscom}
\end{table}

\begin{table}
\begin{centering}
\caption{Statistical properties of Anscombe's quartet data in Table \ref{tab:origAnscom}.}
\par\end{centering}
\begin{centering}
\begin{tabular}{cll}
\hline 
Index & Property & Value\tabularnewline
\hline 
1 & The sample size & $N=11$\tabularnewline
2 & The mean of $x$ & $\bar{x}=9.0$\tabularnewline
3 & The variance of $x$ & $\sigma_{x}^{2}=11.00$\tabularnewline
4 & The mean of $y$ & $\bar{y}=7.5$\tabularnewline
5 & The variance of $y$ & $\sigma_{y}^{2}=4.125$\tabularnewline
6 & The slope & $\beta_{1}=0.5$\tabularnewline
\hline 
7 & The $y-$intercept & $\beta_{0}=3.0$\tabularnewline
8 & The coefficient of determination & $R^{2}=0.667$\tabularnewline
\hline 
\end{tabular}
\par\end{centering}
\centering{}\label{tab:PropAnscom}
\end{table}
\begin{table}
\caption{Anscombe's quartet data sorted by $x$ (for datasets I--III), followed
by $y$ (for dataset IV).}
\begin{centering}
\begin{tabular}{|c|r|r|r|r|r|r|}
\hline 
 &  & \multicolumn{1}{r|}{I} & II & III & \multicolumn{2}{r|}{IV}\tabularnewline
\hline 
Index & $x$ & $y$ & $y$ & $y$ & $x$ & $y$\tabularnewline
\hline 
1 & 4.0 & 4.26 & 3.10 & 5.39 & 8.0 & 5.25\tabularnewline
\hline 
2 & 5.0 & 5.68 & 4.74 & 5.73 & 8.0 & 5.56\tabularnewline
\hline 
3 & 6.0 & 7.24 & 6.13 & 6.08 & 8.0 & 5.76\tabularnewline
\hline 
4 & 7.0 & 4.82 & 7.26 & 6.42 & 8.0 & 6.58\tabularnewline
\hline 
5 & 8.0 & 6.95 & 8.14 & 6.77 & 8.0 & 6.89\tabularnewline
\hline 
6 & 9.0 & 8.81 & 8.77 & 7.11 & 8.0 & 7.04\tabularnewline
\hline 
7 & 10.0 & 8.04 & 9.14 & 7.46 & 8.0 & 7.71\tabularnewline
\hline 
8 & 11.0 & 8.33 & 9.26 & 7.81 & 8.0 & 7.91\tabularnewline
\hline 
9 & 12.0 & 10.84 & 9.13 & 8.15 & 8.0 & 8.47\tabularnewline
\hline 
10 & 13.0 & 7.58 & 8.74 & 12.74 & 8.0 & 8.84\tabularnewline
\hline 
11 & 14.0 & 9.96 & 8.10 & 8.84 & 19.0 & 12.50\tabularnewline
\hline 
\end{tabular}
\par\end{centering}
\label{tab:SortedAnscom}
\end{table}

\subsection{Constraints Applied}
The 11 components of $\bm{x}$ in dataset I--III can be represented
as $x_{k}=x_{k-1}+1$ for $k=1, 2, \cdots, N$ with $x_{0}=3$, so that the $k^{\mathrm{th}}$
component is described as $x_{k}=3+k$, and the mean of $\bm{x}$
is calculated as 
\begin{equation}
\overline{x}  =3+\tfrac{1}{2}\left(N+1\right)\label{eq:mean-x-dx1}
\end{equation}
that provides $\bar{x}=9$ for $N=11$. Now, one can use a more flexible
relationship between $x_{k}$ and $x_{k+1}$ by having an arbitrary
interval $a$, such as 
\begin{equation}
x_{k}=x_{k-1}+a\quad\text{for}\quad k=1, 2, \cdots, N
\end{equation}
 and then the two parameters of $a$ and $x_{0}$ can be determined
by preset constraints of $\bar{x}$ and $\sigma_{x}^{2}$, such as
\begin{eqnarray}
a &=&\sigma_{x}\sqrt{\frac{6}{Nm}}\label{eq:res_gap_a}\\
x_{0} &=&\bar{x}-am
\label{eq:res_init_x}
\end{eqnarray}
where $m=\tfrac{1}{2}\left(N+1\right)$ is an mid-point index. Here,
we restrict ourselves to odd $N$ cases for simplicity. Substitution
of $N=11$, $\sigma_{x}=\sqrt{11}$ (obtained from $S_{xx}=110$),
and $\bar{x}=9$ into Eqs. (\ref{eq:res_gap_a}) and (\ref{eq:res_init_x})
results in $a=1$ and $x_{0}=3$, as shown in Table \ref{tab:SortedAnscom}.
On the other hand, dataset IV has a special set of $\bm{x}$, containing
only two values, denoted as $x_{a}\,\left(=x_{1}=\cdots=x_{N-1}\right)$
and $x_{b}\,\left(=x_{N}\right)$. Because the sequential indices
of $x_{a}$ do not influence any statistical analysis, the mean of
$\bm{x}$ is written as 
\begin{equation}
\bar{x}=\frac{\left(N-1\right)x_{a}+x_{b}}{N}
\end{equation}
and further 
\begin{equation}
\left(N-1\right)\delta x_{a}+\delta x_{b}=0\label{eq:delta-x12}
\end{equation}
where $\delta x_{j}=x_{j}-\bar{x}$ for $j=a,b$. Anscombe used the
fixed value of $S_{xx}$, which is represented below, using $\delta x_{a}$
and $\delta x_{b}$, as 
\begin{equation}
S_{xx}=\left(N-1\right)\delta x_{a}^{2}+\delta x_{b}^{2}=\left(N-1\right)N\delta x_{b}^{2}\label{eq:delta-s12-sqr}
\end{equation}
using Eq. (\ref{eq:delta-x12}).  Finally, we obtain  (for $N=11$)
\begin{eqnarray}
\left(x_{a},x_{b}\right) & =&\left(9\pm1,9\mp10\right)=\left(-1,10\right)\,\mbox{or}\,\left(8,19\right)
\end{eqnarray}
where the latter case of $\left(x_{a},x_{b}\right)=\left(8,19\right)$
was chosen in Anscombe's original work \citep{anscombe_graphs_1973}.

When a paired dataset $\left\{ \left(x_{k},y_{k}\right)\right\} _{k=1}^{N}$
is fitted on a straight line, the goodness of the linear regression
is often estimated using the coefficient of determination $R^{2}$
of Eq. (\ref{eq:defR2}). Alternatively, the slope coefficient $\beta_{1}$
can be set as the last constraint, in addition to $\bar{y}$ and $\sigma_{y}^{2}$,
requiring the minimum sample size of  $N=3$ to fully implement
the six statistical constrains. In the next section, we discuss how
to generate the three $x-y$ data points that hold the six statistical
constrains.

\subsection{A minimum data set of three components}

\begin{table}
\caption{Data of three points satisfying Anscombe's statistical restriction.
Note that $\delta x_{3}=-\delta x_{1}$ and $\delta y_{k}^{\left(2\right)}=-\delta y_{k}^{\left(1\right)}$ for $k=1$ to $3$.}
\begin{centering}
\begin{tabular}{|c|>{\raggedleft}p{0.6in}|>{\raggedleft}p{0.6in}|>{\raggedleft}p{0.6in}|}
\hline 
$k$ & $x_{k}$ & $y_{k}^{\left(1\right)}$ & $y_{k}^{\left(2\right)}$\tabularnewline
\hline 
\hline 
1 & 5.6834 & 6.5187 & 5.1647\tabularnewline
\hline 
2 & 9.0000 & 6.1460 & 8.8540\tabularnewline
\hline 
3 & 12.3166 & 9.8353 & 8.4813\tabularnewline
\hline 
\end{tabular}
\par\end{centering}
\label{tab:Anscome-3pts}
\end{table}
Let's consider three consecutive values of $x$, with the predetermined
constraints of $\bar{x}=9$ and $\sigma_{x}^{2}=11$, to have 
\begin{equation}
x_{k}=x_{0}+a\cdot k\quad\text{for}\quad k=-1,0,1
\end{equation}
where $x_{0}=\bar{x}=9$, $a=\sqrt{11}=3.3166$, so that   $\delta x_{1}=-3.3166=-\delta x_{3}$
and $\delta x_{2}=0$. The following equations are obtained for $\bm{y}$
of three components, such as
\begin{eqnarray}
\delta y_{1}+\delta y_{2}+\delta y_{3} & =0\label{eq:sum-del-y}\\
\delta y_{1}^{2}+\delta y_{2}^{2}+\delta y_{3}^{2} & =2\sigma_{y}^{2}\label{eq:sum-del-y-sqr}\\
\delta x_{1}\cdot\left(\delta y_{1}-\delta y_{3}\right) & =2\beta_{1}\sigma_{x}^{2}\label{eq:sum-del-x-del-y}
\end{eqnarray}
using $\delta x_{2}=0$ and $\delta x_{3}=-\delta x_{1}$. Analytic
solutions of $\delta y_{k}$ for $k=1$ to $3$ are obtained as  
\begin{eqnarray}
\delta y_{2} & =\pm\frac{2}{\sqrt{3}}\sqrt{\sigma_{y}^{2}-B_{1}^{2}}\\
\delta y_{1} & =-\frac{1}{2}\delta y_{2}-B_{1}\\
\delta y_{3} & =-\frac{1}{2}\delta y_{2}+B_{1}
\end{eqnarray}
where $B_{1}=\beta_{1}\sigma_{x}^{2}/\delta x_{3}=\frac{\sqrt{11}}{2}$.

\begin{figure}
\begin{centering}
\includegraphics[width=3.25in]{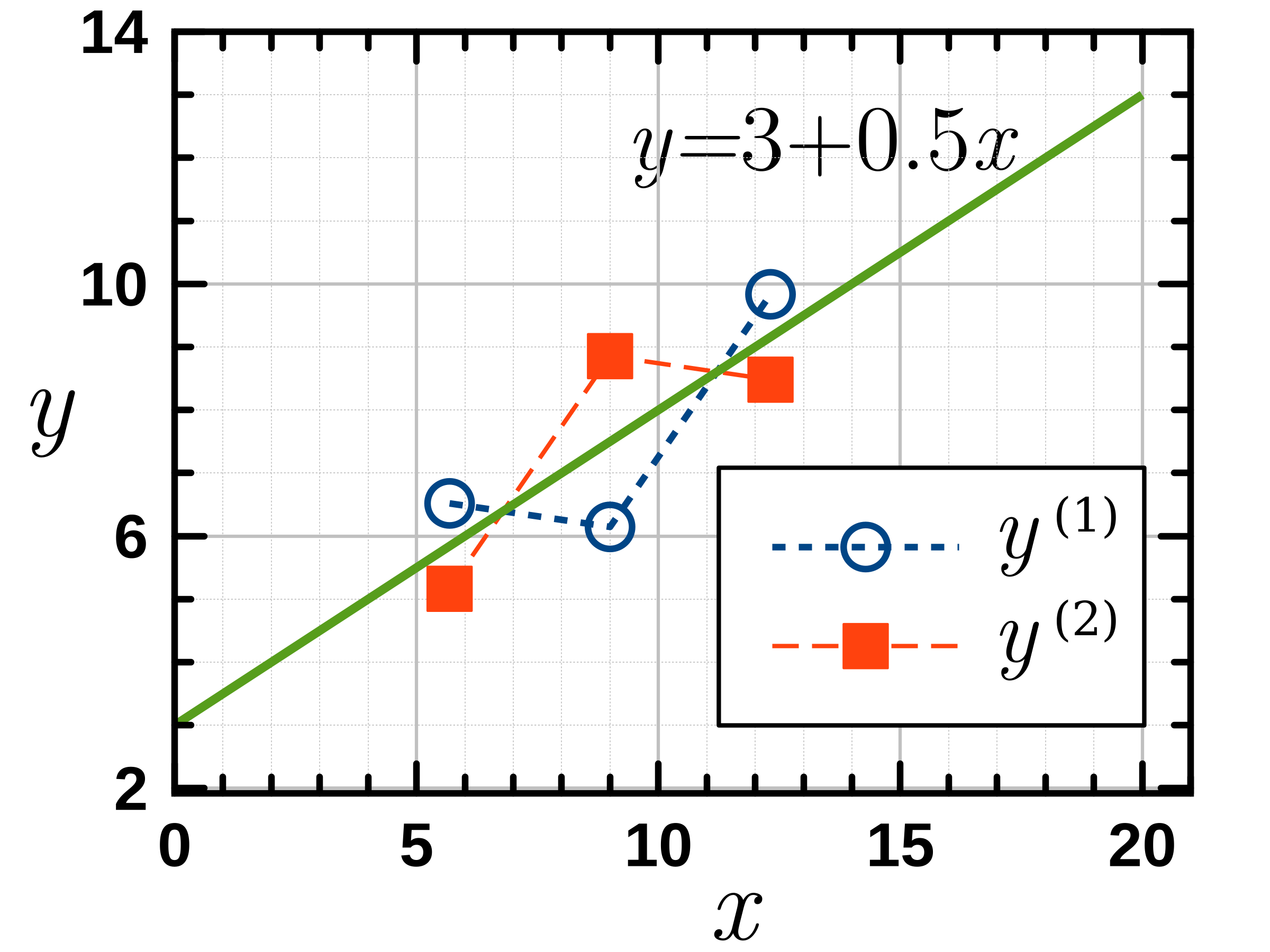}
\par\end{centering}
\caption{Three points satisfying the given linear regression, and $x$ and
$y$ means and variances.}
\label{fig:Anscome-w-3pts}
\end{figure}

Table \ref{tab:Anscome-3pts} shows two sets of solutions, denoted
as $y_{k}^{\left(1\right)}$ and $y_{k}^{\left(2\right)}$, while
both satisfy all of the constraints indicated above. This degeneracy
is due to the squared feature of the variance of Eq. (\ref{eq:sum-del-y-sqr}).
Fig. \ref{fig:Anscome-w-3pts} shows the two sets of $\bm{y}$ versus
$\bm{x}$ with $N=3$, following the same Anscombe's constraints,
except for the sample size. Even with this smallest number of the
sample size for a linear regression, two possible cases of degenerate
$\bm{y}$'s co-exist, having the identical statistical properties.
A trivial case is that if $\sigma_{y}^{2}=B_{1}^{2}$, then $\delta y_{2}=0$
and $\delta y_{1}=-\delta y_{3}=-\beta_{1}\delta x_{3}$, so that
$y_{k}^{\left(1\right)}$ and $y_{k}^{\left(2\right)}$ become identical,
and so will be located on the regression line. 

\subsection{Generation of degenerate datasets with constraints}

\subsubsection{Satisfying three constraints using three arbitrary points}
After the $\bm{x}$ vector of a size of $N$ is determined with constraints
of $\bar{x}$ and $\sigma_{x}^{2}$, other three constraints should
be satisfied by $\bm{y}$ vector, which include finite $\bar{y}$
and $\sigma_{y}^{2}$ , alternately represented as
\begin{equation}
\sum_{k=1}^{N}\delta y_{k} =0\label{eq:sum_dy_zero}
\end{equation}
and 
\begin{equation}
\sum_{k=1}^{N}\delta^{2}y_{k}=S_{yy}=\left(N-1\right)\sigma_{y}^{2}\label{eq:sum_dy2_syy}
\end{equation}
respectively; and finally, $\beta_{1}$ defined as a ratio of a covariance
between $\bm{x}$ and $\bm{y}$ to a variance of $\bm{x}$, i.e.,
$\sigma_{xy}/\sigma_{x}^{2}$, such as 
\begin{equation}
\sum_{k=1}^{N}\delta x_{k}\cdot\delta y_{k}=\left(N-1\right)\sigma_{xy}=\left(N-1\right)\sigma_{x}^{2}\beta_{1}\label{eq:sum_dxdy_syy}
\end{equation}
Because the $\bm{x}$ vector is generated independently, all the constraints
for an arbitrary $N$ are satisfied by the creation of the $\bm{y}$
vector, having a degree of freedom of $N-3$. In our approach, we
generate an initial $\bm{y}$ vector (as a function of the $\bm{x}$
vector), having a specific pattern near the preset regression line
of Eq. (\ref{eq:LR}). Then we select the minimum, maximum, and mid-point
of the $\bm{x}$ vector and adjust the three values of the corresponding
$y$-components to satisfy the constraints of Eqs. (\ref{eq:sum_dy_zero})--(\ref{eq:sum_dxdy_syy}).
 Assume that we already have a sorted $\bm{x}$ vector, i.e., $x_{k-1}<x_{k}$
for $k=1-N$, and have decided $y_{k}$, except $k=1$, $m$, and
$N$, where $m$ is theoretically any index between 1 and $N$, i.e.,
$2\le m\le N-1$. For simplicity, an index of the mid-point can be
used, such as 
\begin{equation}
m=\frac{N+\text{mod\text{\ensuremath{\left(N,2\right)}}}}{2}=\begin{cases}
\tfrac{1}{2}N & \mbox{if }N=\mbox{even}\\
\tfrac{1}{2}\left(N+1\right) & \mbox{if }N=\mbox{odd}
\end{cases}
\end{equation}
where $\text{mod}\left(N,2\right)$ is a remainder when $N$ is divided
by 2, or simply
\begin{equation}
m=\mbox{floor}\left[\tfrac{1}{2}\left(N+1\right)\right]
\end{equation}
which is to round off $\tfrac{1}{2}\left(N+1\right)$, especially for odd $N$.
 The above three equations can be rewritten as 
\begin{eqnarray}
\delta y_{1}+\delta y_{N} & =&-\xsum\delta y_{k}\label{eq:sum_dy_zero-1}\\
\delta y_{1}^{2}+\delta y_{N}^{2} & = &-\xsum\delta^{2}y_{k}+S_{yy}\label{eq:sum_dy2_syy-1}\\
\delta x_{1}\delta y_{1}+\delta x_{N}\delta y_{N} & =&-\xsum\delta x_{k}\delta y_{k}+S_{xx}\beta_{1}\label{eq:sum_dxdy_syy-1}
\end{eqnarray}
where $\xsum=\sum_{k=2}^{N-1}$ is defined as a summation over $k$,
except the first and last indices. Combining Eqs. (\ref{eq:sum_dy_zero-1})
and (\ref{eq:sum_dxdy_syy-1}), we represent $\delta y_{1}$ and $\delta y_{N}$
as linear functions of $\delta y_{m}$, such as
\begin{eqnarray}
\delta y_{1} & =a_{1}+b_{1}\delta y_{m}\label{eq:dy1}\\
\delta y_{N} & =a_{N}+b_{N}\delta y_{m}\label{eq:dyN}
\end{eqnarray}
where 
\begin{eqnarray}
a_{1} & =&\frac{\beta_{1}\left(N-1\right)\sigma_{x}^{2}+\msum\left(\delta x_{n}-\delta x_{k}\right)\cdot\delta y_{k}}{\delta x_{1}-\delta x_{N}}\label{eq:a1}\\
a_{N} & =&a_{1}\label{eq:aN}\\
b_{1} & =&\frac{\delta x_{N}-\delta x_{m}}{\delta x_{1}-\delta x_{N}}\label{eq:b1}\\
b_{N} & =&\frac{\delta x_{1}-\delta x_{m}}{\delta x_{N}-\delta x_{1}}\label{eq:bN}
\end{eqnarray}
and substitute Eqs. (\ref{eq:dy1}) and (\ref{eq:dyN}) into (\ref{eq:sum_dy2_syy-1})
to derive for $\delta y_{m}$, such as 
\begin{equation}
\delta y_{m}=-B\pm\sqrt{s_{yy}^{\prime}+B^{2}-C^{2}}\label{eq:dym}
\end{equation}
where 
\begin{eqnarray}
s_{yy}^{\prime} & =&\left[\left(N-1\right)\sigma_{y}^{2}-\xsum\delta^{2}y_{k}\right]/\left(1+b_{1}^{2}+b_{N}^{2}\right)\\
B & =&\left(a_{1}b_{1}+a_{N}b_{N}\right)/\left(1+b_{1}^{2}+b_{N}^{2}\right)\\
C^{2} & =&\left(a_{1}^{2}+a_{N}^{2}\right)/\left(1+b_{1}^{2}+b_{N}^{2}\right)
\end{eqnarray}
In this case, two sets of $\left\{ \delta y_{1},\delta y_{m},\delta y_{N}\right\} $,
and hence $\left\{ y_{1},y_{m},y_{N}\right\} $, are generated, depending
on the sign of the square-root term in Eq. (\ref{eq:dym}).  Furthermore,
there are no mandatory conditions that the first and last points should
be included to meet the constraints. Instead, three arbitrary points
within a dataset $\left\{ x_{k},y_{k}\right\} _{k=1}^{N}$, e.g.,
$k=p_{1}$, $p_{2}$, and $p_{3}$, can be selected as long as they
are different, i.e., $p_{1}\ne p_{2}\ne p_{3}$. Nevertheless, if
the $x$-positions of the three points are closely located, then large
differences between their $y$-values are expected.

\subsubsection{Selection of a shape function\label{subsec:shape-function}}
In previous sections, we discussed how to determine the components
of the $\bm{y}$ vector, especially three points, assuming that the
rest $N-3$ points are already properly located near the given regression
line of Eq. (\ref{eq:LR}). The predetermination of $N-3$ points
is to have the equal numbers of constraints (3) and unknown points
(3). Because $N-3$ is also the degree of freedom of the pre-positioned
$\bm{y}$ values, the number of patterns that the $N-3$ elements
of $\bm{y}$ can make is theoretically infinite. In addition, having
$\sigma_{y}^{2}$ as one of the constraints doubles the degeneracy
of the created dataset. Here, we consider a function that determines
the initial distribution of $N$ points, among which $y_{1}$, $y_{m}$,
and $y_{N}$ are updated to meet the three statistical constraints
($\bar{y}$, $\sigma_{y}^{2}$ , and $\beta_{1}$). We name  this
function as a shape function and discuss  how shape functions are
used in Anscombe's datasets in the following.

\paragraph*{Random distribution}

In Fig. \ref{fig:Plots-of-Anscombe-Quartet}(a) for dataset I, five
points (of index 2, 5, 7, 8, and 11 in Table \ref{tab:SortedAnscom})
 are located very close to the regression line of $Y$: among the
rest, half of them are above the regression line, and the other half
are below it. The subset consisting of the closest five pairs $\left(x_{j},y_{j}\right)$
of $j=$ 2, 5, 7, 8, and 11 has a regression line of $3.235+0.4746x$
with $R^{2}=0.9950$.  In this case, the shape function of dataset
I is a linear line, similar to the predetermined regression line plus
random biases, such as 
\begin{equation}
f_{I}\left(x_{k}\right)=\tilde{Y}\left(x_{k}\right)+\eta_{k}\left(0,s\right)\label{eq:shape-I}
\end{equation}
where $\tilde{Y}\simeq Y$ and $\bm{\eta}$ is a random vector, having
normally distributed random components with zero mean and a finite
variance, denoted as $s^{2}$. For dataset I, $\bm{\eta}$ should
consist of 11 components, selected from a population of normally distributed
random numbers, with a mean of $\mu\simeq0$ and standard deviation
of $s\simeq\sqrt{1.376}$.

\begin{figure}
\begin{centering}
\includegraphics[width=3.25in]{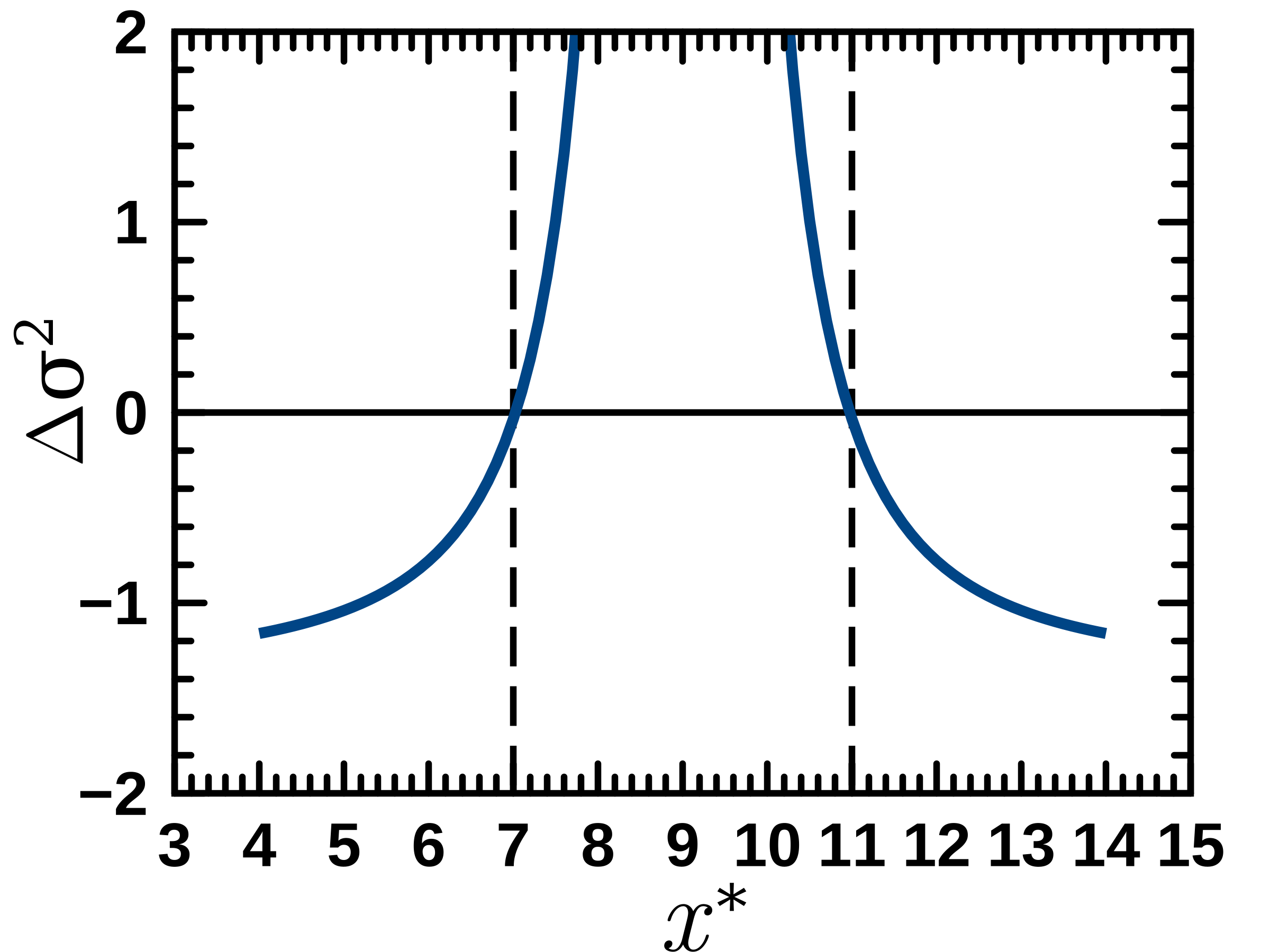}
\par\end{centering}
\caption{The difference between the calculated and predetermined variance of
$y$, i.e., $\Delta\sigma^{2}\left(=\sigma_{y}^{*2}-\sigma_{y}^{2}\right)$
as a function of $x^{*}$. Two values of $x^{*}$ for $\Delta\sigma^{2}=0$
are found as approximately 7 and 11.}
\label{fig:diff_y_var_xstar}
\end{figure}

\paragraph{Quadratic function}
In Fig. \ref{fig:Plots-of-Anscombe-Quartet}(b), the parabolic pattern
of dataset II is best fitted using a quadratic shape function 
\begin{equation}
f_{II}\left(x\right)=q\left(x\right)=q_{0}+\alpha\left(x-x^{*}\right)^{2}\label{eq:shape-II}
\end{equation}
where $q_{0}$ is an $x$-position at an extrema of $f_{II}\left(x=x^{*}\right)\equiv y^{*}$
and $\alpha$ is a coefficient of the quadratic term. From Table \ref{tab:SortedAnscom},
it is straightforward to find that $\left(x^{*},y^{*}\right)=\left(x_{8},y_{8}\right)=\left(11,9.26\right)$.
By trial and error, we found that $\alpha=-0.1256$ with $R^{2}=99.99\%$.
In addition, the flipped (degenerate) shape function is obtained,
such as 
\begin{equation}
f_{II}^{*}\left(x\right)  =2Y\left(x\right)-q\left(x\right)=5.74+0.126\left(x-7\right)^{2}
\end{equation}
and plotted using star symbols. In this case, the three constants
of $q_{0}$, $\alpha$, and $x^{*}$ should be simultaneously determined
to satisfy the three constraints. Eqs. (\ref{eq:sum_dy_zero}), (\ref{eq:sum_dy2_syy}),
and (\ref{eq:sum_dxdy_syy}) are satisfied as follows:
\begin{eqnarray}
\alpha & = & \frac{\beta_{1}\left(N-1\right)\sigma_{x}^{2}}{\sum_{k=1}^{N}\left(x_{k}-x^{*}\right)^{3}}\label{eq:alpha_xstar}\\
q_{0} & = & \bar{y}-\frac{\alpha}{N}\sum_{k=1}^{N}\left(x_{k}-x^{*}\right)^{2}\label{eq:q0_xstar}\\
\sigma_{y}^{2} & = & \frac{\alpha^{2}}{N-1}\sum_{k=1}^{N}\left(x_{k}-x^{*}\right)^{4}-\frac{N\left(q_{0}-\bar{y}\right)^{2}}{N-1}\equiv\sigma_{y}^{*2}\label{eq:vary_xstar}
\end{eqnarray}
indicating that $\alpha=\alpha\left(x^{*}\right)$, $q_{0}=q_{0}\left(\alpha,x^{*}\right)$,
and $\sigma_{y}^{*2}$ as a function of $\alpha$, $q$, and $x^{*}$
with the predetermined constraint $\sigma_{y}^{2}$. Therefore, $x^{*}$
can be obtained by plotting 
\begin{equation}
\Delta\sigma^{2}=\sigma_{y}^{*2}-\sigma_{y}^{2}
\end{equation}
with respect to $x^{*}$ and graphically finding $x^{*}$ of $\Delta\sigma^{2}=0$,
as shown in Fig. \ref{fig:diff_y_var_xstar}. Here, $\alpha=-0.1267$
and $q_{0}=9.2616$ are calculated using Eqs. (\ref{eq:alpha_xstar})
and (\ref{eq:q0_xstar}), respectively, using visually found $x^{*}=10.972$.
Similarly, for $x^{*}=7.027$, we obtained $\alpha=0.1267$ and $q_{0}=5.7405$.
These two parameter sets of $\alpha$, $q_{0}$, and $x^{*}$ are
used to plot $f_{II}$ and $f_{II}^{*}$, shown in Fig. \ref{fig:Plots-of-Anscombe-Quartet}(b).

In Fig. \ref{fig:Plots-of-Anscombe-Quartet}(c), a subset of 10 points,
excluding the outlier of $\left(x_{10},y_{10}\right)=\left(13,12.74\right)$,
are aligned on a straight line, of which the regression line is calculated
as
\begin{equation}
f=\beta_{0}^{\prime}+\beta_{1}^{\prime}x\to f_{III}\label{eq:shape-III}
\end{equation}
where $\beta_{0}^{\prime}=4.01$,  $\beta_{1}^{\prime}=0.3454$, and $R^2=0.999$, 
which can be considered as a shape function of dataset III. Compared
to the given regression line of Eq. (\ref{eq:LR}), $f_{III}$ has
a higher intercept and a gentler slope, as compared to those of Eq.
(\ref{eq:LR}), as well as $\tilde{Y}$ of dataset I. A condition
can be suggested, such as $\left(\beta_{0}^{\prime}-\beta_{0}\right)\left(\beta_{1}^{\prime}-\beta_{1}\right)<0$,
so that if the slope $\beta_{1}^{\prime}$ is stiffer than $\beta_{1}$,
i.e., $\left(\beta_{1}^{\prime}-\beta_{1}\right)>0$; then the intercept
$\beta_{0}^{\prime}$ is located below $\beta_{0}$, and vice versa.
After locating $N$ points on or near the linear shape function of
Eq. (\ref{eq:shape-III}), one arbitrary point, such as $\left(x_{p},y_{p}\right)$
for $p=10$ in dataset III, can be made as an outlier by changing
the $y_{p}$ value. In theory, relocating an outlier position cannot
fully satisfy the three constraints. Instead, this outlier can be
included as one of the three points used for the degenerate dataset's
creation by replacing the mid-point, i.e., $\left\{ \delta y_{1},\delta y_{m},\delta y_{N}\right\} \to\left\{ \delta y_{1},\delta y_{p},\delta y_{N}\right\} $.
Furthermore, the first and last points can also be replaced by any
two distinct points, if needed. 

In Fig. \ref{fig:Plots-of-Anscombe-Quartet}(d) of dataset IV, a group
of ten points is located at a same $x$-positions, i.e., $x_{1}=\cdots=x_{10}=8$.
These ten points have a mean of 7.0 and a variance of 1.527, which
increased to the preset values of $\bar{x}=7.5$ and $\sigma_{x}^{2}=4.125$,
respectively, by including the outlier of $\left(x_{11},y_{11}\right)=\left(19,12.5\right)$
that already satisfies $y_{11}=3+0.5x_{11}$. At $x=8$, $\left\{ y_{k}\right\} _{k=1}^{10}$
can be modeled as normally distributed random numbers of zero mean
and finite variance $s$, such as $\eta\left(0,s\right)$, similar
to Eq. (\ref{eq:shape-I}), such as 
\begin{equation}
f_{IV}\left(x_{k}\right)=\begin{cases}
Y\left(x_{8}\right)+\eta_{k}\left(0,s\right) & \mbox{if }k\ne11\\
3+0.5x_{11} & \mbox{if }k=11
\end{cases}
\end{equation}
Because the last point $\left(x_{11},y_{11}\right)$ is fixed, any
three points at $x=8$ should be (randomly) selected and updated to
meet the given constraints. 

\subsubsection{General algorithm}

If a paired dataset shows a monotonous variation of $y$ with respect
to $x$, or vice versa, then the above-mentioned algorithms can be
generalized and used to create degenerate datasets of the same constrains,
as follows
\begin{enumerate}
\item determine six statistical parameters: $N$, $\overline{x}$, $\sigma_{x}^{2}$,
$\overline{y}$, $\sigma_{y}^{2}$ and $\beta_{1}$ and calculate
$\beta_{0}=\bar{y}-\beta_{1}\bar{x}$;
\item make an $\bm{x}$ vector of $N$ components, having predetermined
$\overline{x}$ and $\sigma_{x}^{2}$;
\item use a shape function to initialize $y$-components near the given
regression line, $Y\left(x\right)$, of Eq. (\ref{eq:LR});
\item update the $y$-values of any three points, as needed, using Eqs.
(\ref{eq:dy1}), (\ref{eq:dyN}), and (\ref{eq:dym}) to satisfy the
three constraints of $\overline{y}$, $\sigma_{y}^{2}$ and $\beta_{1}$;
\item confirm that the generated dataset is characterized by the six parameters
listed above.
\end{enumerate}
In general, when $n_{C}\left(\ge3\right)$ constraints are implemented,
$N-n_{C}$ points are determined using a shape function, and the rest
of the $n_{C}$ points can be determined by analytically solving the
constraint equations. The availability of analytic solutions will
be limited as the number of constraints increases. In this case, the
problem can be described as a linear regression with constraint functions
of 
\begin{equation}
g_{\alpha}\left(y_{1}^{\dagger},...,y_{\beta}^{\dagger},...,y_{n_c}^{\dagger}\right)=0\quad\mbox{for}\quad\alpha,\beta=1-n_{C}
\end{equation}
where $y_{\beta}^{\dagger}$ is one of $n_{C}$ components of $\bm{y}$,
chosen to satisfy $n_{C}$ constraints. A general root-finding algorithm can be 
used to numerically find $y^{\dagger}$.

\section{Results and Discussions}

\subsection{Inverse Sampling of Degenerate Datasets}

The creation of one of degenerate paired datasets requires six constraints,
such as $N$, $\bar{x}$, $\sigma_{x}^{2}$, $\bar{y}$, $\sigma_{y}^{2}$
and $\beta_{1}$. For a given sample size $N$, the mean and standard
deviation of $\bm{x}$ and $\bm{y}$ determine their central locations
and spread degrees. If more than three constraints are considered,
the degrees of degeneracy are theoretically infinite, and therefore,
one can create as many as degenerate datasets as needed, disregarding
their graphical similarities or dissimilarities. If a trend line is
made by a linear regression of multiple datasets of the same size,
then it is highly probable that the
degenerate datasets created from the calculated trend line do not
include the original datasets used for the linear regression, but
instead can include unexpected forms of meaningful datasets.

In statistical physics, the importance sampling technique is frequently
used for efficient Monte Carlo simulations \citep{allen_computer_2017,chen_monte_2006},
which indicates sampling from only specific distributions that over-weigh
the important region. In a microcanonical ensemble of a thermodynamic
system, sampling of particles' positions and velocities are under a constant total 
energy  \citep{schranz_efficient_1991}. Once particle positions
are determined, a specific value of kinetic energy $K$ is calculated as
the total energy subtracted by the position-dependent potential energy,
such as $K=\tfrac{1}{2}\sum_{i}m_{i}v_{i}^{2}$, and particle velocities
are randomly assigned and carefully adjusted to maintain the kinetic
energy. There are many distinct configurations of velocities in the
phase space that give the same kinetic energy value. This specific
sampling is called inverse sampling, which is analogous to the present
work that inversely calculates and samples datasets of specific statistical
constraints.

\subsection{Paired datasets generated using shape functions}

\begin{figure}
\begin{centering}
\includegraphics[width=5in]{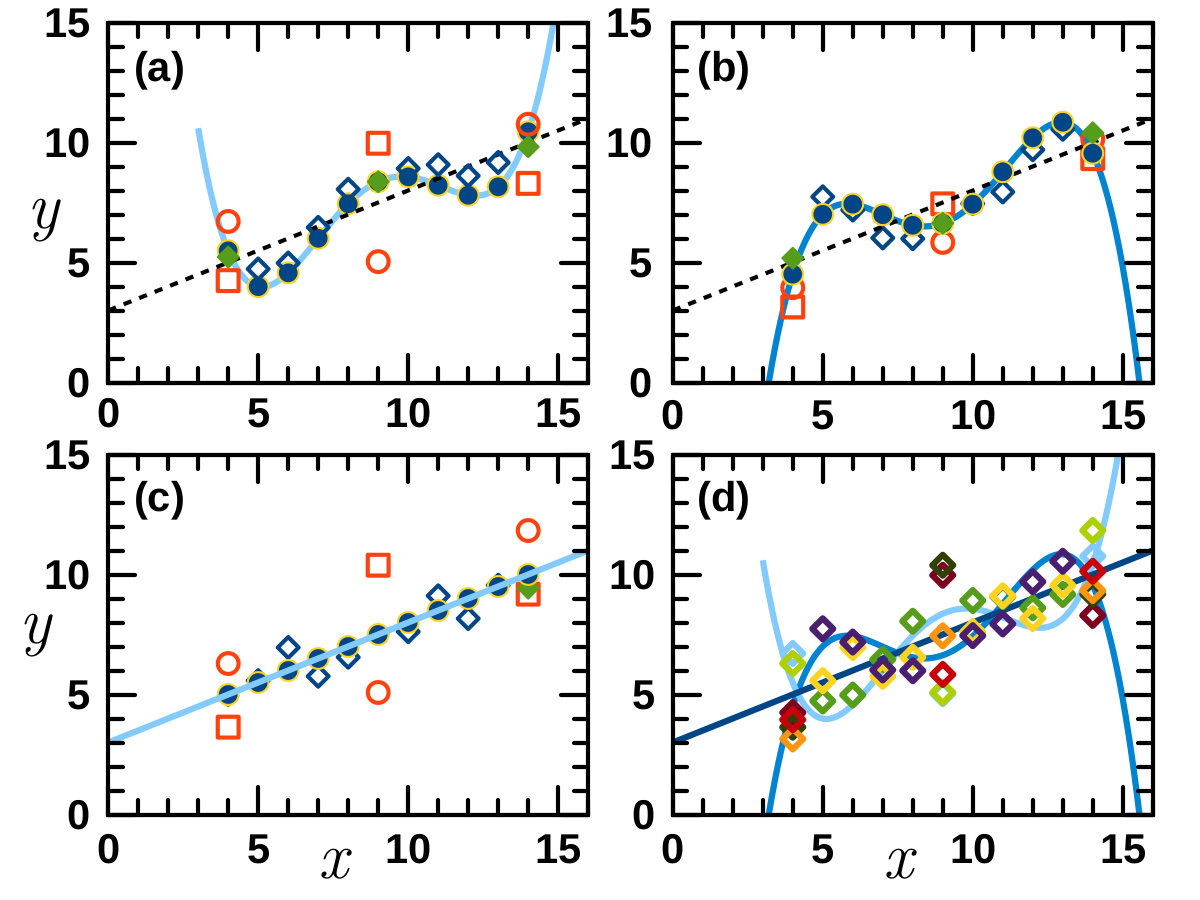}
\par\end{centering}
\caption{Created data using the shape function of Eq. (\ref{eq:shape4th})
with (a) $f_{0}=+\sqrt{2}\times10^{-2}$, (b) $f_{0}=-\sqrt{2}\times10^{-2}$,
and (c) $f_{0}=0$: filled circles are points on the shape function;
blank diamonds are randomly deviated from the shape functions; three
filled diamonds are replaced by either hollow squares or circles,
adjusted to meet the statistical constraints listed in Table \ref{tab:PropAnscom};
and (d) a collection of all hollow symbols in (a)--(c), where specific
patterns become unnoticeable.}
\label{fig:degen-poly4}
\end{figure}

In this section, we create a few paired datasets having the same statistical
properties of Anscombe's quartet. A test shape function we employed
is a fourth-order polynomial with respect to $x$, such as
\begin{equation}
f\left(x\right)=Y\left(x\right)+f_{0}(x-h_{1})(x-h_{2})(x-h_{3})(x-h_{4})\label{eq:shape4th}
\end{equation}
where $h_{1}=4.150$, $h_{2}=7.480$, $h_{3}=10.710$, and $h_{4}=13.850$
are chosen slightly away from the (integer) $x_{k}$ value; and $f_{0}$
is a weight factor of the shape function, which is either 0 or $\pm\sqrt{2}\times10^{-2}$.
The non-zero magnitude of $f_{0}$ is selected by trial and error.
If $f_{0}=0$.0, then the predetermined regression line $Y\left(x\right)$
becomes the shape function $f\left(x\right)$ itself. Linear regressions
using the shape function are summarized in Fig. \ref{fig:degen-poly4},
as discussed below.

\begin{enumerate}
\item These shape functions of positive, negative, and zero $f_{0}$ values
are made and shown in Fig. \ref{fig:degen-poly4}(a), (b), and (c),
respectively. Figs. \ref{fig:degen-poly4}(a) and (b) show smooth
shape functions with opposite signs, and Fig. \ref{fig:degen-poly4}(c)
shows the predetermined linear regression line $Y\left(x\right)$
as the shape function so that initial points are located on the regression
line. Fig. \ref{fig:degen-poly4}(d) combines all the datasets and
shape functions, generated in (a)--(c), and shows the overall trend.
\item In each of Figs. \ref{fig:degen-poly4}(a)--(c), $N$ points initially
on the shape function (filled circles) are randomly relocated to new
neighboring positions (hollow diamonds) above or below the original
positions.
\item Three vertical positions of $y_{j}$ for $j=1$, $m$, and $N$ are
adjusted to two distinct groups (hollow circles and rectangles), of
which both satisfy the given constraints using the algorithm discussed
above.
\end{enumerate}

Since there are, in principle, an infinite number of available shape
functions and multiple ways to locate data points near the shape
functionss, identifying data patterns seems to be challenging
work, even if the trend seems to follow a noticeable shape. 
Nevertheless, this work provides in-depth analysis of Anscombe's
original work in terms of data creation, and a straightforward algorithm
to reproduce his work, as well as to perform the inverse sampling
of regressible datasets.

\onecolumngrid 

\twocolumngrid

\subsection{Further Considerations}

While $\bm{x}$ is predetermined independently, the three effective constraints
provide the same number of closure equations, reducing the degrees
of freedom of $\bm{y}$ from $N$ to $N-3$. It is worth noting that
the degeneracy is originated not only by the number of elements, but
also from the squared form of $\sigma_{y}^{2}=E\left[\left(y-\bar{y}\right)^{2}\right]$,
i.e., the expected value of $\left(y-\bar{y}\right)^{2}$ or the second
central moment. Even if we have only three points ($N=3$), two datasets
are generated due to the intrinsic degenerate characteristics of the
variance, as previously shown in Fig. \ref{fig:Anscome-w-3pts}. 
After $N-3$ points are decided for a paired sample of
$N$, there are still two degrees of degeneracy, due to the squared
feature of variances.

In addition to the three constraints discussed above, one can include
more constraints, such as, but not limited to, the least magnitude
(instead of squared) of errors, linear regression with perpendicular
offsets \citep{sampaio_iterative_2006}, heteroscedasticity \citep{breusch_simple_1979},
Kolmogorov-Smirnov statistic\citep{massey_kolmogorov-smirnov_1951},
Lagrange multiplier (LM) statistic \citep{breusch_simple_1979}, standardized
skewness, standardized Kurtosis, and D-statistic\citep{cook_detection_1977}.
Each of these statistics can be used as an additional constraint to
quantitatively identify statistical similarities. Adding or replacing
some of the above-mentioned constraints to the standard constraints
will create too many distinct datasets to visually compare. However,
if the created datasets are tested for various indices, they can easily
be classified in to several groups of similarities  \citep{sala-i-martin_i_1997}.

Table \ref{tab:moment34} shows the third and fourth moments of the
$z$-scores of $x$ and $y$, denoted as $z_{x}=\left(z-\bar{x}\right)/\sigma_{x}$
and $z_{y}=\left(y-\bar{y}\right)/\sigma_{y}$, respectively. The
$n^{\text{th }}$ moment of $z$ is defined as a mean value of $z^{n}$,
i.e., $\left\langle z^{n}\right\rangle $, and the third and fourth
moments are called standard skewness and kurtosis, respectively. Because
the $\bm{x}$'s of datasets I--III are equally evenly distributed,
their mean, variance, skewness, and kurtosis are identical, which
is not observed in dataset IV. The $y$-skewness of dataset I and
II have negative values, indicating that more $y$-values are located
below $\bar{y}$. The larger magnitude of $\left\langle z_{y}^{3}\right\rangle _{II}=-0.97882$
than $\left\langle z_{y}^{3}\right\rangle _{I}=-0.04837$ indicates
the quadratic shape function of dataset II locates more data points
lower than the regression line than those of dataset I. The two largest
$y$-skewnesses of dataset III and IV are ascribed to their outliers.
The $x$-kurtosis values of the four datasets show a similar trend
to those of skewnesses, and the $y$-Kurtosis values increase from
dataset I to dataset IV, also following the similar trend of absolute
$y$-skewnesses. Including higher order moments will require the same
number of additional constraints to make multiple datasets statistically
identical within the range of constraints applied.

\begin{table}
\caption{The third and fourth moments of Anscombe's quartet.}
\begin{centering}
\begin{tabular}{|>{\centering}p{0.5in}|>{\raggedleft}p{0.5in}|>{\raggedleft}p{0.5in}|>{\raggedleft}p{0.5in}|>{\raggedleft}p{0.5in}|}
\hline 
Dataset & $\left\langle z_{x}^{3}\right\rangle $ & $\left\langle z_{x}^{4}\right\rangle $ & $\left\langle z_{y}^{3}\right\rangle $ & $\left\langle z_{y}^{4}\right\rangle $\tabularnewline
\hline 
\hline 
I & 0.000 & 1.471 & $-$0.048 & 1.801\tabularnewline
\hline 
II & 0.000 & 1.471 & $-$0.979 & 2.486\tabularnewline
\hline 
III & 0.000 & 1.471 & 1.377 & 4.228\tabularnewline
\hline 
IV & 2.467 & 7.521 & 1.119 & 3.622\tabularnewline
\hline 
\end{tabular}
\par\end{centering}
\label{tab:moment34}
\end{table}
\section{Conclusion}
\label{sec:conc} 
Testing the similarities or identicalness of two datasets from distinct
origins is an ubiquitously important issue in statistics, applicable
to various studies. When a paired dataset is linearly regressed, the
trend line indicates correlation degrees of how the response variable
$y$ depends on the independent variable $x$, assuming that one is
a cause and the other is an effect, or vice versa.  In reality, it
is rare to have two or more visually different datasets that provide
an identical regression equation. On the other hand, a given regression
equation can interpret or explain a number of datasets from various
sources. Here, we recognized that a robust method to sample many degenerate
datasets satisfying the given constraints is of great necessity, not
only in advanced data sciences and applications, but also in applied
statistics education at college and graduate levels. In this work,
we presented an algorithm to sample many degenerate datasets having
the identical six constraints used for a linear regression. Our method
is extendable for an arbitrary number of constraints, including higher-order
statistical moments, to create statistically closer datasets.
 

%

\end{document}